\documentclass[a4paper,twoside,10pt]{article}

\input{jgrg19.sty}
\begin{document}
%
\pagestyle{fancy}
\fancyhead{}
  \fancyhead[RO,LE]{\thepage}
  \fancyhead[LO]{K.~Nakamura}                  
  \fancyhead[RE]{
    Consistency of Equations for the Single Scalar Field Case in 
    Second-order ...
  }    
\rfoot{}
\cfoot{}
\lfoot{}
\label{P7}    
\title{%
  Consistency of Equations for the Single Scalar Field Case in 
  Second-order Gauge-invariant Cosmological Perturbation Theory
}
%
\author{%
  Kouji Nakamura\footnote{Email address: kouji.nakamura@nao.ac.jp}
}
%
\address{%
  Optical and Infrared Astronomy Division, National
  Astronomical Observatory of Japan., Osawa, Mitaka, Tokyo,
  181-8588, Japan.
}
%
\abstract{
  We derived the second-order perturbations of the Einstein
  equations and the Klein-Gordon equation for a generic
  situation in terms of gauge-invariant variables.
  The consistency of all the equations is confirmed.
  This confirmation implies that all the derived equations of
  the second order are self-consistent and these equations are
  correct in this sense.
  We also discuss the physical implication of these
  equations.
}

\section{Introduction}
\label{sec:intro}


The general relativistic second-order cosmological perturbation
theory is one of topical subjects in the recent cosmology. 
Recently, the first-order approximation of our universe from a
homogeneous isotropic one was revealed through the observation
of the CMB by the Wilkinson Microwave Anisotropy Probe
(WMAP)\cite{WMAP}, the cosmological parameters are accurately
measured, we have obtained the standard cosmological model, and
the so-called ``precision  cosmology'' has begun.
These developments in observations were also supported by the
theoretical sophistication of the linear order cosmological
perturbation theory.
To explore more detail observations, the Planck satellite was
launched on the last May and its first light was
reported\cite{Planck}.
With the increase of precision of the CMB data, the study of
relativistic cosmological perturbations beyond linear order is a
topical subject.
The second-order cosmological perturbation theory is one
of such perturbation theories beyond linear order.


In this article, we show a part of our formulation of the
second-order gauge-invariant perturbation
theory\cite{kouchan-second-series}.
We give the consistency relations of the source terms in all the
second-order perturbation of the Einstein equations and the
Klein-Gordon equation in the single scalar field case as in the
case of the perfect fluid case\cite{Nakamura:2009sa}.
These consistency relations imply the all derived equations of
the second order are self-consistent and these equations are
correct in this sense.
Further, we also discuss the physical implication of our
second-order Einstein equations.


\section{Metric and matter perturbations}
\label{sec:Second-order-cosmological-perturbatios}


The background spacetime for the cosmological perturbations is a
homogeneous isotropic background spacetime.
The background metric is given by
\begin{eqnarray}
  \label{eq:background-metric}
  g_{ab} = a^{2}\left\{
    - (d\eta)_{a}(d\eta)_{b}
    + \gamma_{ij} (dx^{i})_{a} (dx^{j})_{b}
  \right\},
\end{eqnarray}
where $\gamma_{ab} := \gamma_{ij} (dx^{i})_{a} (dx^{j})_{b}$ is
the metric on the maximally symmetric three-space and the
indices $i,j,k,...$ for the spatial components run from 1 to 3.
On this background spacetime, we consider the perturbative
expansion of the metric as
$\bar{g}_{ab} = g_{ab} + \lambda {}_{{\cal X}}\!h_{ab}
+ \frac{\lambda^{2}}{2} {}_{{\cal X}}\!l_{ab}
+ O(\lambda^{3})$,
where $\lambda$ is the infinitesimal parameter for perturbation
and $h_{ab}$ and $l_{ab}$ are the first- and the second-order
metric perturbations, respectively. 
As shown in Refs.~\cite{kouchan-second-series}, the metric
perturbations $h_{ab}$ and $l_{ab}$ are decomposed as 
\begin{eqnarray}
  h_{ab} =: {\cal H}_{ab} + {\pounds}_{X}g_{ab},
  \label{eq:linear-metric-decomp}
  \quad
  l_{ab}
  =:
  {\cal L}_{ab} + 2 {\pounds}_{X} h_{ab}
  + \left(
      {\pounds}_{Y}
    - {\pounds}_{X}^{2}
  \right)
  g_{ab},
\end{eqnarray}
where ${\cal H}_{ab}$ and ${\cal L}_{ab}$ are the
gauge-invariant parts of $h_{ab}$ and $l_{ab}$, respectively.
The components of ${\cal H}_{ab}$ and ${\cal L}_{ab}$ can be
chosen so that 
\begin{eqnarray}
  \label{eq:components-calHab}
  {\cal H}_{ab}
  &=& 
  a^{2} \left\{
    - 2 \stackrel{(1)}{\Phi} (d\eta)_{a}(d\eta)_{b}
    + 2 \stackrel{(1)}{\nu}_{i} (d\eta)_{(a}(dx^{i})_{b)}
    + \left( - 2 \stackrel{(1)}{\Psi} \gamma_{ij} 
      + \stackrel{(1)}{\chi}_{ij} \right)
    (dx^{i})_{a}(dx^{j})_{b}
  \right\},
  \\
  \label{eq:components-calLab}
  {\cal L}_{ab}
  &=& 
  a^{2} \left\{
    - 2 \stackrel{(2)}{\Phi} (d\eta)_{a}(d\eta)_{b}
    + 2 \stackrel{(2)}{\nu}_{i} (d\eta)_{(a}(dx^{i})_{b)}
    + \left( - 2 \stackrel{(2)}{\Psi} \gamma_{ij} 
      + \stackrel{(2)}{\chi}_{ij} \right)
    (dx^{i})_{a}(dx^{j})_{b}
  \right\}.
\end{eqnarray}
In Eqs.~(\ref{eq:components-calHab}) and
(\ref{eq:components-calLab}), the vector-mode
$\stackrel{(p)}{\nu}_{i}$ and the tensor-mode
$\stackrel{(p)}{\chi_{ij}}$ ($p=1,2$) satisfy the properties 
\begin{eqnarray}
  D^{i}\stackrel{(p)}{\nu}_{i} =
  \gamma^{ij}D_{p}\stackrel{(p)}{\nu}_{j} = 0, \quad
  \stackrel{(p)}{\chi^{i}_{\;\;i}} = 0, \quad
   D^{i}\stackrel{(p)}{\chi}_{ij} = 0,
\end{eqnarray}
where $\gamma^{kj}$ is the inverse of the metric $\gamma_{ij}$.


On the other hand, we also expand the scalar field as
$\bar{\varphi} = \varphi + \lambda \hat{\varphi}_{1} +
\frac{\lambda^{2}}{2} \hat{\varphi}_{2} + O(\lambda^{3})$ and
decompose $\hat{\varphi}_{1}$ and $\hat{\varphi}_{2}$ into
gauge-invariant and gauge-variant parts as
\begin{eqnarray}
  \hat{\varphi}_{1} =: \varphi_{1} + {\pounds}_{X}\varphi, \quad
  \hat{\varphi}_{2} =: \varphi_{2} + 2 {\pounds}_{X}\varphi_{1}
  + \left({\pounds}_{Y} - {\pounds}_{X}^{2}\right)\varphi,
\end{eqnarray}
respectively, where $X^{a}$ and $Y^{a}$ are the gauge-variant
parts of the first- and the second-order metric perturbations,
respectively, in Eqs.~(\ref{eq:linear-metric-decomp}).


\section{Equations for Perturbations}
\label{sec:Cosmological-Einstein-equations-Klein-Gordon-equations}


Here, we summarize the Einstein equations and the Klein-Gordon
equations for the background, the first order, and the second
order on the above background spacetime (\ref{eq:background-metric}).


The background Einstein equations for a single scalar field
system are given by 
\begin{eqnarray}
  \label{eq:background-Einstein-equations-scalar-1}
  &&
  {\cal H}^{2} + K = \frac{8 \pi G}{3} a^{2} \left(
    \frac{1}{2a^{2}} (\partial_{\eta}\varphi)^{2} + V(\varphi)
  \right)
  ,\quad
  2 \partial_{\eta}{\cal H} + {\cal H}^{2} + K = 8 \pi G 
  \left(-\frac{1}{2} (\partial_{\eta}\varphi)^{2} + a^{2} V(\varphi)\right)
  ,
\end{eqnarray}
where ${\cal H}:=\partial_{\eta}a/a$, $K$ is the curvature
constant of the maximally symmetric three-space.


On the other hand, the second-order perturbations of the
Einstein equation are summarized as 
\begin{eqnarray}
  &&
    2 \partial_{\eta} \stackrel{(2)}{\Psi}
  + 2 {\cal H} \stackrel{(2)}{\Phi}
  -
  8\pi G \varphi_{2} \partial_{\eta}\varphi  
  = 
  \Delta^{-1} D^{k} \Gamma_{k}
  , \quad
  \stackrel{(2)}{\Psi} - \stackrel{(2)}{\Phi}
  = 
  \frac{3}{2}
  (\Delta + 3 K)^{-1}
  \left\{
    \Delta^{-1} D^{i}D_{j}\Gamma_{i}^{\;\;j} - \frac{1}{3} \Gamma_{k}^{\;\;k}
  \right\}
  \label{eq:kouchan-18.213}
  , \\
  && 
  \left\{
    \partial_{\eta}^{2}
    + 2 \left(
      {\cal H}
      - \frac{\partial_{\eta}^{2}\varphi}{\partial_{\eta}\varphi}
    \right)
    \partial_{\eta}
    -             \Delta
    -          4  K
    + 2 \left(
      \partial_{\eta}{\cal H}
      - \frac{\partial_{\eta}^{2}\varphi}{\partial_{\eta}\varphi} {\cal H}
    \right)
  \right\}
  \stackrel{(2)}{\Phi}
  \nonumber\\
  && 
  \quad\quad\quad\quad
  =
  - \Gamma_{0}
  - \frac{1}{2} \Gamma_{k}^{\;\;k}
  +
  \Delta^{-1} D^{i}D_{j}\Gamma_{i}^{\;\;j}
  + \left(
    \partial_{\eta}
    - \frac{\partial_{\eta}^{2}\varphi}{\partial_{\eta}\varphi}
  \right)
  \Delta^{-1}D^{k}\Gamma_{k}
  \nonumber\\
  && 
  \quad\quad\quad\quad\quad\quad
  -
  \frac{3}{2}
  \left\{
    \partial_{\eta}^{2}
    - \left(
      \frac{2\partial_{\eta}^{2}\varphi}{\partial_{\eta}\varphi} - {\cal H}
    \right)
    \partial_{\eta}
  \right\}
  (\Delta + 3 K)^{-1}
  \left\{
    \Delta^{-1} D^{i}D_{j}\Gamma_{i}^{\;\;j} - \frac{1}{3} \Gamma_{k}^{\;\;k}
  \right\}
  \label{eq:kouchan-18.233}
  , \\
  &&
  \stackrel{(2)}{\nu_{i}}
  = 
  \frac{2}{\Delta + 2 K}
  \left\{
    D_{i} \Delta^{-1} D^{k} \Gamma_{k}
    - \Gamma_{i}
  \right\}
  ,
  \quad
  \partial_{\eta}
  \left(
    a^{2} \stackrel{(2)}{\nu_{i}}
  \right)
  =
  \frac{2 a^{2}}{\Delta + 2 K}
  \left\{
    D_{i}\Delta^{-1} D^{k}D_{l}\Gamma_{k}^{\;\;l}
    - D_{k}\Gamma_{i}^{\;\;k}
  \right\}
  \label{eq:kouchan-18.199-3}
  , \\
  &&
  \left(
    \partial_{\eta}^{2} + 2 {\cal H} \partial_{\eta} + 2 K  - \Delta
  \right)
  \stackrel{(2)\;\;\;\;}{\chi_{ij}}
  \nonumber\\
  && \quad\quad
  =
  2 \Gamma_{ij}
  - \frac{2}{3} \gamma_{ij} \Gamma_{k}^{\;\;k}
  - 3
  \left(
    D_{i}D_{j} - \frac{1}{3} \gamma_{ij} \Delta
  \right) 
  \left( \Delta + 3 K \right)^{-1}
  \left(
    \Delta^{-1} D^{k}D_{l}\Gamma_{k}^{\;\;l}
    - \frac{1}{3} \Gamma_{k}^{\;\;k}
  \right)
  \nonumber\\
  && \quad\quad\quad\quad
  + 4
  \left\{ 
      D_{(i} (\Delta+2K)^{-1} D_{j)}\Delta^{-1}D^{l}D_{k}\Gamma_{l}^{\;\;k}
    - D_{(i}(\Delta+2K)^{-1}D^{k}\Gamma_{j)k}
  \right\}
  = 
  0
  ,
\end{eqnarray}
where $\Gamma_{i}^{\;\;j} := \gamma^{kj}\Gamma_{ik}$ and
$\Gamma_{k}^{\;\;k} = \gamma^{ij}\Gamma_{ij}$.
The source terms $\Gamma_{0}$, $\Gamma_{i}$, and $\Gamma_{ij}$
are the collections of the quadratic terms of the linear-order
perturbations in the second-order Einstein equations.
Further, the second-order perturbation of the Klein-Gordon
equation 
\begin{eqnarray}
  &&
       \partial_{\eta}^{2}\varphi_{2} 
  +  2 {\cal H} \partial_{\eta}\varphi_{2} 
  -    \Delta\varphi_{2} 
  -    \left(
    \partial_{\eta}\stackrel{(2)}{\Phi}
    +  3 \partial_{\eta}\stackrel{(2)}{\Psi}
  \right) \partial_{\eta}\varphi
  +  2 a^{2} \stackrel{(2)}{\Phi} \frac{\partial V}{\partial\bar{\varphi}}(\varphi)
  +    a^{2}\varphi_{2}\frac{\partial^{2}V}{\partial\bar{\varphi}^{2}}(\varphi)
  = \Xi_{(K)}
  ,
  \label{eq:Klein-Gordon-eq-second-gauge-inv-explicit}
\end{eqnarray}
where the source term $\Xi_{(K)}$ is also the collections of the
quadratic terms of the linear-order perturbations in the
second-order Klein-Gordon equation.
The explicit form of these $\Gamma_{0}$, $\Gamma_{i}$,
$\Gamma_{ij}$, and $\Xi_{(K)}$ are given in
Refs.~\cite{kouchan-second-series}.
The first-order perturbations of the Einstein equations are
given by the replacements
$\stackrel{(2)}{\Phi}\rightarrow\stackrel{(1)}{\Phi}$, 
$\stackrel{(2)}{\Psi}\rightarrow\stackrel{(1)}{\Psi}$, 
$\stackrel{(2)\;\;}{\nu_{i}}\rightarrow\stackrel{(1)\;\;}{\nu_{i}}$, 
$\stackrel{(2)\;\;\;\;}{\chi_{ij}}\rightarrow\stackrel{(1)\;\;\;\;}{\chi_{ij}}$, 
$\varphi_{2}\rightarrow\varphi_{1}$, and
$\Gamma_{0}=\Gamma_{i}=\Gamma_{ij}=\Xi_{(K)}=0$.


\section{Consistency of equations for second-order perturbations}
\label{sec:Consistency-of-the-second-order-perturbations}


Now, we consider the consistency of the second-order
perturbations of the Einstein equations
(\ref{eq:kouchan-18.213}) and (\ref{eq:kouchan-18.233}) for 
the scalar modes, Eqs.~(\ref{eq:kouchan-18.199-3}) for vector
mode, and the Klein-Gordon equation
(\ref{eq:Klein-Gordon-eq-second-gauge-inv-explicit}).


Since the first equation in Eqs.~(\ref{eq:kouchan-18.199-3}) is
the initial value constraint for the vector mode
$\stackrel{(2)}{\nu_{i}}$ and it should be consistent with the
evolution equation, i.e., the second equation of
Eqs.~(\ref{eq:kouchan-18.199-3}).
Explicitly, these equations are consistent with each other if
the equation
\begin{eqnarray}
  \partial_{\eta}\Gamma_{k}
  + 2 {\cal H} \Gamma_{k}
  - D^{l}\Gamma_{lk} = 0
  \label{eq:kouchan-19.358}
\end{eqnarray}
is satisfied.
Actually, through the first-order perturbative Einstein
equations, we can directly confirm the equation
(\ref{eq:kouchan-19.358}) through the background Einstein
equations, the first-order Einstein equations, and the long
expressions of $\Gamma_{i}$ and $\Gamma_{ij}$ given in
Refs.\cite{kouchan-second-series}.


Next, we consider the consistency of the second-order
perturbation of the Klein-Gordon equation
(\ref{eq:Klein-Gordon-eq-second-gauge-inv-explicit}) and the
Einstein equations (\ref{eq:kouchan-18.213}) and
(\ref{eq:kouchan-18.233}).
From these equation, we can show that the second-order
perturbation of the Klein-Gordon equation is consistent with the
background and the second-order Einstein equations if the
equation 
\begin{eqnarray}
  &&
  2 \left(
    \partial_{\eta} + {\cal H}
  \right) \Gamma_{0}
  -    D^{k}\Gamma_{k}
  +    {\cal H} \Gamma_{k}^{\;\;k}
  + 8 \pi G \partial_{\eta}\varphi \Xi_{(K)}
  = 0
  \label{eq:kouchan-19.374}
\end{eqnarray}
is satisfied under the background and the first-order Einstein 
equations.
Further, we can directly confirm Eq.~(\ref{eq:kouchan-19.374})
through the background Einstein equations, the first-order
perturbation of the Einstein equations, and the long expression
of $\Gamma_{0}$, $\Gamma_{i}$, $\Gamma_{ij}$, and $\Xi_{(K)}$
which are given in Refs.~\cite{kouchan-second-series}.


Equation (\ref{eq:kouchan-19.358}) comes from the consistency of
the initial value constraint and evolution equation and
Eq.~(\ref{eq:kouchan-19.374}) comes from the consistency between
the Klein-Gordon equation and the Einstein equation.
These equation should be trivially satisfied from a general
viewpoint, because the Einstein equation is the first class
constrained system.
However, these trivial results imply that we have derived the
source terms $\Gamma_{0}$, $\Gamma_{i}$, $\Gamma_{ij}$, and
$\Xi_{(K)}$ are consistent with each other and are correct in 
this sense.
We also note that these relations are independent of the
details of the potential of the scalar field.


\section{Summary and discussions}
\label{sec:summary}


In this article, we summarized the second-order Einstein
equation for a single scalar field system.
We derived all the components of the second-order perturbation
of the Einstein equation without ignoring any types modes
(scalar-, vector-, tensor-types) of perturbations.
As in the case of the perfect fluid\cite{Nakamura:2009sa}, we
derived the consistency relation between the source terms of the
second-order Einstein equation and the Klein-Gordon equation.


In our formulation, any gauge fixing is not necessary and we
can obtain all equations in the gauge-invariant form, which are
equivalent to the complete gauge fixing.
In other words, our formulation gives complete gauge-fixed
equations without any gauge fixing.
Therefore, equations obtained in a gauge-invariant manner cannot
be reduced without physical restrictions any more.
In this sense, the equations shown here are irreducible.
This is one of the advantages of the gauge-invariant perturbation
theory.


The resulting Einstein equations of the second order show that
the mode-couplings between different types of modes appears as
the quadratic terms of the linear-order perturbations owing to
the nonlinear effect of the Einstein equations, in principle.
Perturbations in cosmological situations are classified into
three types: scalar, vector, and tensor.
In the second-order perturbations, we also have these three
types of perturbations as in the case of the first-order
perturbations.
In the scalar field system shown in this article, the
first-order vector mode does not appear due to the momentum
constraint of the first-order perturbation of the Einstein
equation.
Therefore, we have seen that three types of mode-coupling
appear in the second-order Einstein equations, i.e.,
scalar-scalar, scalar-tensor, and tensor-tensor type of mode
coupling.
Since the tensor mode of the linear order is also generated due
to quantum fluctuations during the inflationary phase, the 
mode-couplings of the scalar-tensor and tensor-tensor types may
appear in the inflation.
If these mode-couplings occur during the inflationary phase,
these effects will depend on the scalar-tensor ratio $r$.
If so, there is a possibility that the accurate observations of
the second-order effects in the fluctuations of the scalar type
in our universe also restrict the scalar-tensor ratio $r$ or
give some consistency relations between the other observations
such as the measurements of the B-mode of the polarization of
CMB.
This will be a new effect that gives some information on the
scalar-tensor ratio $r$.


As the current status of the second-order gauge-invariant
cosmological perturbation theory, we may say that the curvature
terms in the second-order Einstein tensor, i.e., the
second-order perturbations of the Einstein tensor, are almost
completely derived, although there remain some problems should
be clarified\cite{kouchan-second-series}.
The next task is to clarify the nature of the second-order
perturbation of the energy-momentum tensor through the extension
to multi-fluid or multi-field systems.
Further, we also have to extend our arguments to the Einstein
Boltzmann system to discuss CMB physics, since we have to treat
photon and neutrinos through the Boltzmann distribution
functions. 
This issue is also discussed in some
literature\cite{Non-Gaussianity-in-CMB}. 
If we accomplish these extension, we will be able to clarify the
Non-linear effects in CMB physics.



\end{document}